# Thin-film Lithium Niobate on Insulator Surface Acoustic Wave Devices for 6G Centimeter Bands

Tzu-Hsuan Hsu[1,2], Joshua Campbell[1], Jack Kramer[1], Sinwoo Cho[1], Zhi-Qiang Lee[2], Ming-Huang Li[2] and Ruochen Lu[1]
[1]*The University of Texas at Austin,* Austin, TX, U.S.A.
[2]*National Tsing Hua University,* Hsinchu, Taiwan
tzuhsuan.hsu@austin.utexas.edu

*Abstract*—In this work, we investigate the frequency scaling of shear-horizontal (S.H.) surface acoustic wave (SAW) resonators based on a lithium niobate on insulator (LNOI) substrate into the centimeter bands for 6G wireless systems. Prototyped resonators with wavelengths ranging between 240 nm and 400 nm were fabricated, and the experimental results exhibit a successful frequency scaling between 9.05 and 13.37 GHz. However, a noticeable performance degradation can be observed as the resonance frequency ($f_s$) scales. Such an effect is expected to be caused by non-ideal $h_{elec}/\lambda$ for smaller $\lambda$ devices. The optimized LNOI SH-SAW with a $\lambda$ of 400 nm exhibits a $f_s$ of 9.05 GHz, a $k_{eff}^2$ of 15%, $Q_{max}$ of 213 and a FoM of 32, which indicates a successful implementation for device targeting centimeter bands.

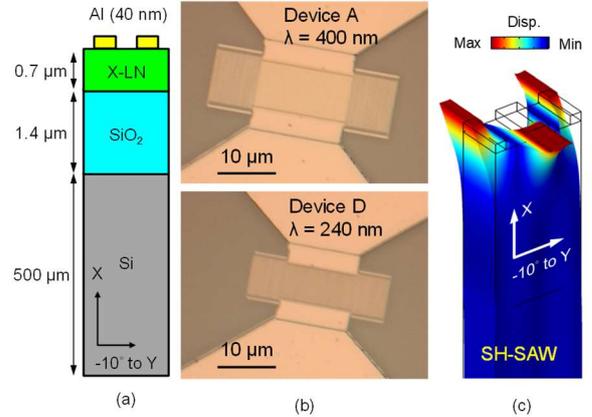

Fig. 1. (a) Unit cell model of the SH-SAW resonator based on X-LN/SiO$_2$/Si. (b) The optical microscope image of the fabricated SH-SAW resonators. (c) Simulated mode shape of the SH-SAW resonator.

## I. INTRODUCTION

With the rapid development of the 5G wireless networks [1], the so-called Frequency Range 1 (FR1) bands in 5G new radio (5G NR) have successfully extended the R.F. band application to sub-6 GHz [2] for urban capacity expansion. In the meantime, the Frequency Range 2 (FR2) bands covering 24-71 GHz millimeter wave (mmWave) [3] serve as the backbone for hot spot applications. Therefore, although the initial call for next generation wireless standard or 6G focuses on further development into the sub-THz range for short-range or sensing applications [4], there have been multiple calls from the industry on extending the FR1 radio into the 7-20 GHz mid-bands (so-called 6G centimeter bands [5]). Such bands can offer larger bandwidth for further urban capacity expansion and are expected to be compatible with existing rooftop infrastructures. However, major obstacles during the implementation of the 6G centimeter bands include the wide variety of the new centimeter band spectra with different contiguous spectrum availability, in addition to the integration of the sub-6 GHz 5G spectrum into the 6G system. As a result, it is expected that in the next-generation radio frequency front-end (RFFE), there will be a need for diverse sets of filter bandwidths and array compositions. Hence, the cost-effect frequency scalability for R.F. acoustic component design will be an important task in the pursuit of next-generation wireless systems at centimeter bands.

Traditionally, RFFE markets have been captured by surface acoustic wave (SAW) devices [6][7] based on bulk piezoelectric materials in the MHz to GHz frequency range and the aluminum nitride (AlN) based bulk acoustic wave (BAW) devices in the higher frequency range from GHz and above [8]. However, following the shift towards emerging bands, novel R.F. acoustic devices are required to achieve higher frequencies, larger bandwidths, and lower losses. These emerging bands have significantly strained the existing technology owing to the limited electromechanical coupling coefficient ($k_{eff}^2$) of conventional acoustic device and the adjacent bandwidth they can provide. Therefore, various research attempts have been proposed to explore the scalability [9] or improvement of the R.F. performance at higher frequencies through the introduction of scandium aluminum nitride (ScAlN) [10]. Yet, the complex manufacturing process for BAW devices and the need for individual trimming may suggest that it might not be the most cost-effective solution.

On the other hand, research efforts are also proposed based on suspended thin-film piezoelectric waveguides such as the Lamb wave resonators (LWRs). The thin-film lithium tantalate (LT) [11][12] or lithium niobate (LN) [13][14] waveguide has been proven to feature high Q or large coupling. Previous results have been presented to scale toward the mmWave range by targeting the higher-order modes [15]. Additionally, examples demonstrated through employing the periodically poled LN have provided a pathway to mitigate the internal charge cancellation and performance loss during frequency scaling [16], and the initial results indicate good performance for R.F. filters [19]. Yet, the need for individual trimming on specific LN thickness in order to reach a certain frequency still limits the design flexibility on single-chip frequency scaling.

On the contrary to BAW devices, the shear-horizontal surface acoustic wave (SH-SAW) devices based on innovative hetero acoustic layered (HAL) [18] wafers for RFFE applications gained attention from both industries [19][20] and research groups [21] recently. Examples include LN on insulator [22] and LT on insulator [23]. These devices not only successfully demonstrated performance improvements in both the $k_{eff}^2$ and Q by forming an acoustic waveguide through acoustic velocity mismatch between layers [24] but also retained the multi-band operation flexibility [25] as the resonance frequency ($f_s$) can still be lithographically defined, just like any SAW device. The solid mount and ease of fabrication also suggest that the HAL SAW resonator can be a cost-effective solution that is worth further investigation. Hence, considering that thin-film LN SH-SAW resonator based on lithium niobate on insulator (LNOI) stacking has previously demonstrated excellent figure-of-merit (FoM = $k_{eff}^2 \times Q$) in the sub-GHz to 2 GHz range [26], the potentials of LNOI SH-SAW scaling toward centimeter bands are studied in this work. The stacking selected consists of LN 0.7 μm, silicon dioxide ($SiO_2$) 1.4 μm, and high-resistive silicon substrate, as depicted in Fig. 1(a).

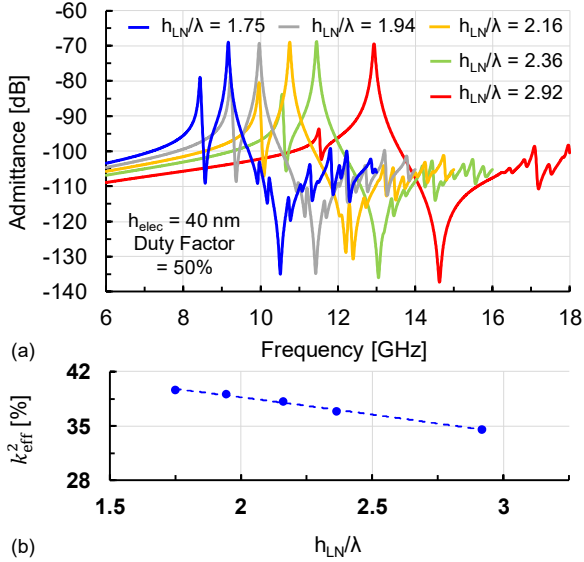

Fig. 2. (a) Simulated admittance spectra of the LNOI SH-SAW resonators. The quality factor ($Q_{set}$) in the simulation is set to be 200. (b) Extracted electromechanical coupling ($k_{eff}^2$) from the admittance plot.

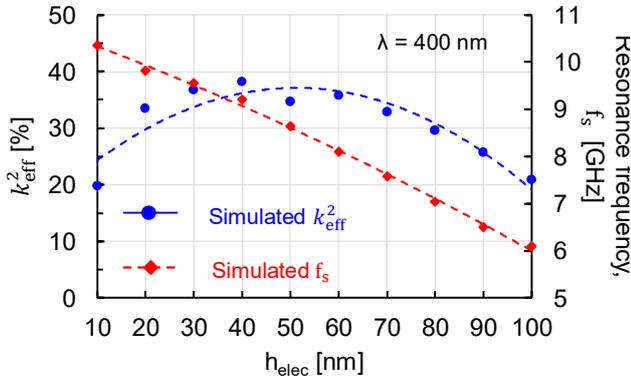

Fig. 3. Simulated $k_{eff}^2$ and resonance frequency ($f_s$) of a SH-SAW resonator with a wavelength (λ) of 400 nm. Strong dispersive behavior of both $k_{eff}^2$ and $f_s$ is observed due to electrode loading.

## II. SH-SAW RESONATOR DESIGN FOR CENTIMETER BANDS

### A. Frequency scaling in thin-film lithium niobate

To study the frequency scaling of SH-SAW resonators toward centimeter bands, the periodic unit cell model depicted in Fig. 1(a) with aluminum (Al) electrodes set at 40 nm is used to simulate the RF performance. The SH-SAW propagation direction is selected by aligning along the -10° to Y-axis of an X-cut LN following the discussion previously made in [27] to maximize the $k_{eff}^2$. The fabricated interdigital transducers (IDTs) and spaced reflecting gratings (RGs) used to stimulate the SH-SAW are shown in Fig. 1(b). The mode shape simulated for the desired SH-SAW is then depicted in Fig. 1(c). The established finite element method (FEM) detailly described in [27] was used to simulate the dispersion effect of different LN thicknesses to wavelengths ratios ($h_{L.N.}/λ$) ranging from 1.75 up to 2.92. The electrode configuration was set to follow the previously mentioned 40 nm of Al and an IDT duty factor of 50%. The quality factor ($Q$) used in all simulations is set to 200, and the results depicted in Fig. 2(a) exhibit a successful frequency scaling ranging between 9.16 GHz and 12.93 GHz. The $k_{eff}^2$ simulated is then extracted using (1) [13] where $f_s$ and $f_p$ represent series and parallel resonance frequency. The extracted $k_{eff}^2$ is summarized in Fig. 2(b).

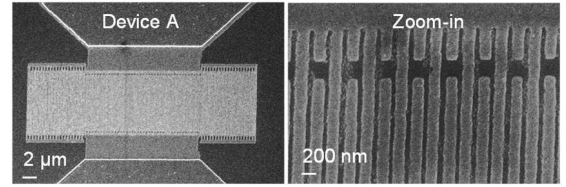

| Device | λ [nm] | $h_{LN}/λ$ | $N_e$ | $N_r$ | A [λ] | Duty Cycle [%] |
|---|---|---|---|---|---|---|
| A | 400 | 1.75 | 40 | 40 | 20 | 70 |
| B | 360 | 1.94 | 40 | 40 | 20 | 50 |
| C | 324 | 2.16 | 54 | 40 | 30 | 50 |
| D | 296 | 2.36 | 42 | 40 | 30 | 50 |
| E | 240 | 2.92 | 64 | 40 | 30 | 50 |
| F | 400 | 1.75 | 40 | 40 | 20 | 50 |

Fig. 4. Summary of different design parameters presented in this work and the scanning electron microscope image of the fabricated LNOI SH-SAW resonators with a minimum critical dimension of 100 nm.

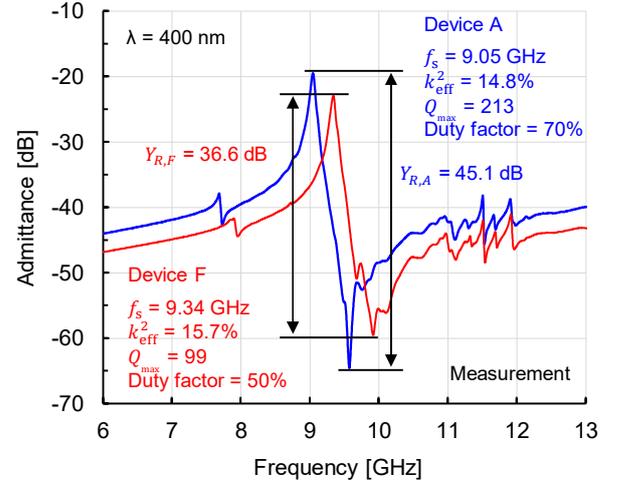

Fig. 5. Measured admittance response of two devices with different IDT duty factors. The admittance ratio ($Y_R$) is improved by implementing a higher duty factor.

$$k_{eff}^2 = \pi^2/8 \cdot (f_p^2 - f_s^2)/f_s^2 \quad (1)$$

The result ranges between 39% to 34% according to different $h_{L.N.}/λ$. Therefore, it indicates a promising opportunity for the LNOI SH-SAW into the centimeter bands.

### B. Effects from different electrode thicknesses

The arrangement of electrode configuration also plays an important role during the design consideration of the centimeter band LNOI SH-SAW. Al is chosen as the electrode material as it has less impact on the phase velocity of the SH-SAW and is relatively beneficial for our design towards higher frequency as compared to other heavier metal such as gold [27] or Copper [28]. Furthermore, owing to the aggressive λ shrinking needed to reach the desired centimeter bands, the electrode thickness ($h_{elec}$) to λ ratio also significantly increases as compared to lower frequency devices [29]. For instance, even though $h_{elec}$ has also been reduced from 300 nm [29] to 40 nm, the $h_{elec}/λ$ still increases from 0.075 to 0.16 since the λ has been shrunk from 4 μm to 0.24 μm. Such increase in $h_{elec}/λ$ may result in the device performance being much more sensitive to slight $h_{elec}$ variation. As depicted in Fig. 3, with a λ set at 400 nm, by varying the $h_{elec}$ between 10 to 100 nm, a significant change in both $k_{eff}^2$ and $f_s$ is observed. As $h_{elec}$ approaches 10 nm, though it can achieve the highest $f_s$, not only does the $k_{eff}^2$ effected by the electrodes being unable to concentrate the strain energy towards the surface, but also suffers from large series resistance ($R_s$). On the other hand, the more $h_{elec}$ there is, the more mass loading effect on $f_s$ is observed. Hence, a $h_{elec}$ of 40 nm is selected by optimizing for both $k_{eff}^2$ and frequency scaling at λ = 400 nm.

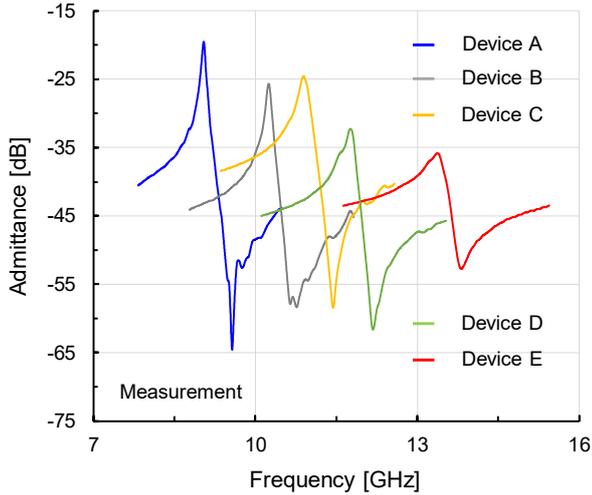

Fig. 6. Measured admittance of Device A to Device E. The effect of electrode loading is clearly visible as the frequency scaled above 10GHz.

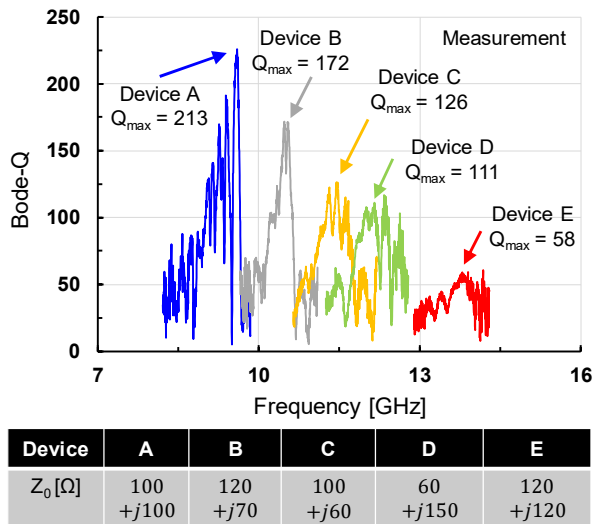

| Device | A | B | C | D | E |
|---|---|---|---|---|---|
| $Z_0$ [Ω] | 100 +j100 | 120 +j70 | 100 +j60 | 60 +j150 | 120 +j120 |

Fig. 7. Extracted Bode-Q plot for different resoantor designs.

## III. MEASUREMENT RESULTS

### A. Fabrication results and performance characteristics

The proposed LNOI SH-SAW resonators were fabricated through electron beam lithography and a metal lift-off process.

TABLE I: Perfomance summary of each resonator presented in this work.

| Device name | λ [nm] | $h_{LN}/λ$ | $f_s$ [GHz] | $k_{eff}^2$ [%] | $Q_{max}$ | FoM = $k_{eff}^2 × Q_{max}$ |
|---|---|---|---|---|---|---|
| A | 400 | 1.75 | 9.05 | 15 | 213 | 32 |
| B | 360 | 1.94 | 10.25 | 11 | 172 | 19 |
| C | 324 | 2.16 | 10.89 | 13 | 126 | 16 |
| D | 296 | 2.36 | 11.77 | 9 | 111 | 10 |
| E | 240 | 2.92 | 13.37 | 7 | 58 | 4 |
| F | 400 | 1.75 | 9.34 | 16 | 99 | 16 |

*$k_{eff}^2$ extracted through mBVD model fitting.

The detailed design parameters, including λ, number of electrodes ($N_e$), number of RGs ($N_r$), aperture (A), and duty factors, are depicted in Fig. 4 atop the scanning electron microscope image taken on device A. The zoom-in view revealed good fabrication results on IDTs.

To verify the performance of the LNOI SH-SAW resonators. The RF response is measured using standard R.F. probing techniques, including the Keysight P5028A VNA and G.S.G. probing. The standard short-open-load-thru (SOLT) calibration process was carefully performed, while no data de-embedding technique was used after the data acquisition. Fig. 5 depicts the measurement result with two resonators featuring the same λ of 400 nm, device A with a higher duty factor of 70% exhibiting a $f_s$ of 9.05 GHz, a $k_{eff}^2$ of 15% and a $Q_{max}$ of 213. On the contrary, device F, with a conventional duty factor of 50% only features a slightly higher $f_s$ of 9.34 GHz and a $Q_{max}$ of 99. The difference is also clearly visible in the admittance ratio ($Y_R$) defined in (2) where $Y_s$ and $Y_p$ are defined as the admittance at series and parallel resonances. Such improvements are expected to be attributed to the ohmic resistive loss reduction caused by the thin IDT fingers.

$$Y_R = Y_s/Y_p \quad (2)$$

### B. Challenges in frequency scaling and electrode loading

To further study the behavior of SH-SAW resonators with aggressively scaled $h_{L.N.}/λ$, the admittance response of device A to E is plotted in Fig. 6, representing $h_{L.N.}/λ$ from 1.75 to 2.92. It is obvious that as the relatively large $h_{elec}/λ$ significantly loaded the performance of the resonator. The Bode-Q and the $Q_{max}$ of each device are carefully extracted through the PathWave Advanced Design System (A.D.S.) software following (3) and the procedure listed in [30][31]. Note that the source impedance corresponding to each device has been tuned to the center of the measured S parameter in the Smith chart. [32]

$$Q_{Bode} = ω · \left|\frac{dS_{11}}{dω}\right| · \frac{1}{1 - |S_{11}|^2} \quad (3)$$

As a result, a clear trend of performance degradation on the extracted Bode-Q and $Q_{max}$ can be observed in Fig. 7. Lastly, the overall performance of all six devices presented in this work is summarized in Table. 1 where the device A exhibits optimal performance merits. This is likely the result of the proper electrode configuration design. Therefore, a decent FoM of 32 can be achieved. A degradation of $k_{eff}^2$ is also observed as the frequency scales, which is likely caused by both electrode mass loading and the ohmic loss from IDTs. However, the large deviation in absolute value in $k_{eff}^2$ between simulation and measurement is also noticed in this work, which may prompt further investigation and understanding of the additional loss mechanisms at this frequency and scale range that may not exist in our existing model.

IV. CONCLUSIONS

In this work, we explore the possibilities of scaling the SH-SAW resonator into the centimeter bands using the established LNOI platform. The significance of proper electrode design is intensively studied and experimentally verified. Promising performance for LNOI SH-SAW resonator operating at 9.05 GHz with λ of 400 nm, a $k_{eff}^2$ of 15%, $Q_{max}$ of 213, and a FoM of 32 is achieved. Yet, the results also indicate the importance of both process control and electrode configuration in this device scale, as any slight variation in the $h_{elec}/\lambda$ may have a significant impact on the resonator. Hence, future design optimization and investigation on the loss mechanisms dominating such bands are needed. Furthermore, it is expected that with the use of substrate with higher phase velocity such as sapphire [33] or silicon carbide [34], the acoustic energy confinement of the acoustic waveguide can be improved. Nonetheless, the LNOI SH-SAW resonator still demonstrates its potential to become a key building block in the next-generation wireless systems.

ACKNOWLEDGMENT

The authors would like to thank the funding support from the DARPA COFFEE program.